\documentclass[a4paper,11pt]{article}
\pdfoutput=1 

\usepackage{jheppub} 

\usepackage[T1]{fontenc} 
\usepackage[latin1]{inputenc}
\usepackage{slashed}
\usepackage{amsmath}
\usepackage{textcomp}
\usepackage{amssymb}
\usepackage{amsfonts}
\usepackage{indentfirst}
\usepackage{color}
\usepackage{hyperref}
\usepackage{subcaption}
\usepackage[shortlabels]{enumitem}
\usepackage[normalem]{ulem}
\usepackage{comment}

\newcommand{\be}{\begin{equation}}
\newcommand{\ee}{\end{equation}}
\newcommand{\bea}{\begin{eqnarray}}
\newcommand{\eea}{\end{eqnarray}}

\usepackage{graphicx,graphics}
\graphicspath{{figures/}}

\begin{document}

\hfill{KCL-PH-TH/2023-05}

\title{\boldmath Cosmic bounce and phantom-like equation of state from tunnelling}

\author[a]{Jean Alexandre,} 
\author[a]{and Silvia Pla}  

\affiliation[a]{Theoretical Particle Physics and Cosmology, King's College London, WC2R 2LS, UK}

\emailAdd{jean.alexandre@kcl.ac.uk}
\emailAdd{silvia.pla\_garcia@kcl.ac.uk}

\abstract{
We allow a scalar field on a flat FLRW background metric to tunnel between two degenerate vacua. The resulting true vacuum state then violates the Null Energy Condition, and the corresponding homogeneous fluid induces a bounce, after which it has a phantom-like equation of state and asymptotically leads to a de Sitter phase. 
The mechanism presented here requires no exotic matter or modified gravity, it is purely generated by quantum fluctuations and is valid for a generic double well potential. }

\maketitle

\flushbottom

\newpage



\section{Introduction}

Unlike spontaneous symmetry breaking (SSB), which occurs in infinite volume, tunnelling involves remarkable energetic features, among which a non-perturbative ground state with no classical analogue. This is at the origin of convexity of the effective potential for a scalar field \cite{Symanzik:1969ek,Coleman:1974jh,Iliopoulos:1974ur,Haymaker:1983xk,Fujimoto:1982tc,Bender:1983nc,Hindmarsh:1985nc,Plascencia:2015pga,Millington:2019nkw}, and thus symmetry restoration.

The explicit calculation of the one-particle-irreducible (1PI) effective potential, taking into account several degenerate vacua, 
was done in \cite{Alexandre:2012ht,Alexandre:2022sho} in the semi-classical approximation for the partition function. These studies assumed an $O(4)$-symmetric Euclidean space-time, and the corresponding work at finite but low temperature was done in \cite{Alexandre:2022qxc} and \cite{Alexandre:2023jsl}. 
The latter works allow for the full tunnelling regime, 
involving a gas of Euclidean-time-dependent instantons relating two degenerate vacua.
It was found that the true ground state for the scalar field is symmetric, and it
violates the Null Energy Condition (NEC - see \cite{Rubakov:2014jja,Kontou:2020bta} for reviews), because it is non-extensive
in the thermodynamical sense. We note here that these results are independent of symmetry-restoration by the 
Kibble-Zurek mechanism \cite{Kibble:1976sj,Zurek:1985qw}, which is valid at high temperatures and does not allow for the NEC to be violated.

The present work extends this tunnelling mechanism to a Friedmann-Lemaitre-Robertson-Walker (FLRW) background metric, 
where we study the backreaction of the fluid provided by the scalar true vacuum on the metric dynamics.
Our assumptions do not involve exotic matter or modified gravity, 
but a finite volume and an adiabatic expansion instead, both to be defined in the next section.
Our results arise purely from quantum fluctuations and they have no classical counterpart.

It is well known that in Quantum Field Theory (QFT) the energy conditions can be violated under certain circumstances. Some examples include the Casimir effect \cite{Casimir:1948}, radiation from moving mirrors \cite{Davies:1977yv}, or black hole evaporation \cite{Hawking:1975vcx}. Another interesting example in curved backgrounds was obtained in \cite{Onemli:2002hr}. The latter work studies a self-interacting massless field, therefore seeing only one vacuum, and not tunnelling. Also, the background is fixed as a de Sitter metric, whereas in our study the scale factor is determined by the backreaction of the scalar effective vacuum. Nevertheless, it is still possible for the stress-energy tensor to satisfy certain constraints, such as the Averaged Null Energy Condition (ANEC),  which averages the NEC over timelike or null geodesics. 
The mechanism we propose here indeed does not violate the ANEC, since NEC violation is valid temporarily only - see Sec.\ref{Sec.Friedmann}. 
We note that an eternal inflation scenario is described in \cite{Kontou:2020pdx}, which also respects the ANEC.

In Sec.\ref{Sec.semiclassical} we describe the semi-classical approximation in which we evaluate the partition function,
based on the different saddle points which are relevant for two degenerate vacua: two static configurations and a gas of 
instantons/anti-instantons.  In the situation of non-degenerate vacua, the relevant configurations are the Coleman bounce \cite{Coleman1,Coleman2}
and the shot \cite{Andreassen:2016cvx}, with imaginary quantum fluctuations which arise from a negative eigenvalue in the fluctuation determinant \cite{Ai:2019fri}.
In the present case though, there are not any imaginary quantum corrections, since the (anti-) instantons are monotonic functions of Euclidean time \cite{Kleinert:2004ev}.

The effective action is then derived in Sec.\ref{Sec.effectiveaction} to the lowest order in the field,
which is enough to confirm convexity and that the ground state is obtained for a vanishing field, unlike the situation of SSB.
This calculation is done in the adiabatic approximation, assuming that the tunnelling rate is large compared to the space-time expansion rate.
The vacuum energy induced by tunnelling violates the NEC, which results in a cosmic bounce if we start with a contracting universe. This is followed by an expansion with a phantom-like equation of state $w<-1$, which asymptotically becomes de Sitter with $w=-1$.
The property $w<-1$ is usually related to a negative kinetic term in the potential 
(see \cite{Ludwick:2017tox} for a review on phantom energy),
but is not the case here: the vacuum we find is homogeneous and its energetic properties  
arise purely from quantum fluctuations, not from a specific bare action.
 We note that the origin of energies is not relevant to this mechanism, 
since the effective ground state energy is always lower than the degenerate bare energies, and it is this negative difference which induced NEC violation.

In Sec.\ref{Sec.Friedmann} we solve numerically the Friedmann equations, where we study the backreaction of the effective theory on gravity. As expected from NEC violation, the solution exhibits a cosmological bounce \cite{Steinhardt:2001st,Khoury:2001wf,Khoury:2001bz}, known to provide an alternative to Cosmic Inflation \cite{Guth:1980zm,Linde:1981mu}. The original idea to generate a bounce from a tunnelling-induced scalar field true vacuum was proposed in \cite{Alexandre:2019ygz,Alexandre:2021imu}, in the context of an $O(4)$-symmetric Euclidean space-time though, whereas we allow here for the full tunnelling regime, with finite volume and infinite Euclidean time.

Finally, the detailed calculations are presented in Appendix A, B and C.

\section{Semi-classical approximation and saddle points}\label{Sec.semiclassical}

\subsection{Assumptions}

We consider the classical background metric 
\be
\text{d}s^2=-\text{d}t^2+a^2(t)\delta_{ij}\text{d}x^i \text{d}x^j~,
\ee
where the scale factor $a(t)$ is kept generic. The bare matter action is 
\be \label{eq:Smatter}
S[\phi]=\int \text{d}^4 x \sqrt{|g|} ( L - j \phi)~,
\ee
where the Lagrangian $L$ involves a double-well potential, 
as well as a non-minimal coupling to the scalar curvature:
\be \label{eq:Lagrangian-1}
L=-\frac{1}{2}g^{\mu \nu}\partial_\mu \phi \partial_\nu \phi - \frac{1}{2}\xi R \phi^2
-\frac{\lambda}{4!}(\phi^2-v^2)^2 -\bar\Lambda~.
\ee 
For convenience, we have also added the cosmological constant term in the matter sector ($\bar \Lambda=\kappa^{-1}\Lambda$ with $\kappa=8\pi G$) to account for vacuum energy effects after renormalisation. 
The important assumptions we make are the following:
\begin{itemize}

    \item Finite volume, which allows tunnelling between the degenerate vacua. We start from 
    a fundamental flat spatial cell with volume $V_0$ and comoving volume $a^3(t)V_0$,
    which can be thought of as a 3-torus, or a 3-sphere with large enough radius to neglect curvature.
    Although finite, we assume the parameters of the model to be such that quantisation of momentum can be ignored,
    and the periodic boundary conditions do not play a role. Related comments on the Casimir effect are given in \cite{Alexandre:2023jsl} for tunnelling in flat space-time, 
    and we focus here on the tunnelling features only;
    
    \item Adiabatic approximation, where the expansion rate of the metric 
    is assumed small compared to the tunnelling rate for matter. 
    According to the discussion at the end of Sec.\ref{Sec.gas}, this is valid in the regime 
    \be\label{validity}
    |H|\equiv\left|\frac{\dot a}{a}\right|\ll v \sqrt\frac{\lambda}{\pi}~\alpha^{3/2}~\exp(-\alpha^3)~,
    \ee 
    where $\alpha^3(t)=a^3(t)S_0/\hbar$ and $S_0$ is the action for an instanton interpolating the two vacua $\pm v$.
    
\end{itemize}
As a consequence of the second point, the scale factor $a(t)$ will be considered constant for the calculation of 
the matter effective theory, and its time dependence will be reinstated when we couple the matter effective theory to gravity.

\subsection{Semi-classical approximation}

We work here in Euclidean signature. In the semi-classical approximation, and focusing only on the matter sector for the reasons explained above, the partition function takes the form 
\be
Z[j]=\int{\cal D}[\phi]\exp(-S[\phi]/\hbar)\simeq \sum_n Z_n[j]~,
\ee
where
\be
Z_n= F_n[j] \exp(-S[\phi_n]/\hbar)\equiv\exp(-\Sigma_n[j]/\hbar)~,
\ee
and $\phi_n$ are the different dominant contributions, the saddle points, which satisfy the equation of motion and minimise the action locally in the space of field configurations. $F_n[j]$ are the fluctuation factors for these saddle points, that we will calculate at one-loop, and $\Sigma_n[j]$ are the corresponding connected graphs generating functionals.

The saddle points $\phi_n$ satisfy then
\be \label{eq:class-eom}
-\frac{1}{\sqrt{g}}\frac{\delta S}{\delta \phi}=j~,
\ee
and since we consider two degenerate minima, a bubble-solution cannot form, since it would have an infinite radius \cite{Coleman1,Coleman2}. 
Hence we focus on homogeneous saddle points only, which can depend on the Euclidean time tough. 
These saddle points obey
\be \label{eq:eom-saddle01}
\ddot\phi+\frac{3 \dot a}{a} \dot\phi-\xi R\phi+\frac{\lambda}{6}v^2 \phi -\frac{\lambda}{6} \phi^3=j~,
\ee
where a dot represents a (Euclidean) time derivative.
In the adiabatic approximation, the scale factor $a$ is assumed constant for the calculation of quantum fluctuations for matter, and we will therefore take $\dot a=0=R$. 
We discuss below the static saddle points and the instanton gas, with their corresponding connected graphs generating functionals $\Sigma_1[j],\Sigma_2[j]$ and $\Sigma_{gas}[j]$ respectively. 

Finally, we are interested in the tunnelling-induced effective potential, such that it is enough to consider a constant source $j$. A spacetime-dependent source is necessary for the 
calculation of the derivative part of the effective action only.

\subsection{Static saddle points} \label{subsec:static-s}

The static saddle points satisfy
\be
v^2\phi - \phi^3=\frac{6j}{\lambda}~,
\ee
which, for $j<j_c\equiv\lambda v^3/(9\sqrt3)$, has two real solutions
\be\label{staticsaddle}
\phi_1(j) =\frac{2 v}{\sqrt{3}} \cos \left(\frac{\pi}{3}-\frac{1}{3} \arccos (j/j_c)\right) ~~~~\mbox{and}~~~~
\phi_2(j) =-\phi_1(-j)~,
\ee 
with the corresponding actions 
\bea
S_1[j]&\equiv& S[\phi_1(j)]=\int \text{d}^4x\sqrt{g}\left(\bar\Lambda+v \, j
-\frac{3}{2 v^2\lambda}~ j^2+\mathcal{O}(j^3) \right) \\
S_2[j]&\equiv& S[\phi_2(j)]=S_1[-j]~.\nonumber
\eea 
The one-loop fluctuation factor for a static saddle point $\phi_n(j)$ is calculated in Appendix \ref{static}, using the Schwinger proper time representation of the propagator. We find for the corresponding renormalised connected graphs generating functional  
\bea\label{Sigman}
\Sigma_n[j]&=&\int \text{d}^4 x \sqrt{g}\Bigg( \bar \Lambda_R +\frac{\lambda_R}{4!}(\phi_n^2-v_R^2)\\
&&\qquad \qquad
+\frac{\hbar \lambda_R^2}{4608\pi^2}\Big(G(\phi_n)
+2(3 \phi_n^2-v_R^2)^2\ln\big((3\phi_n^2/v_R^2-1)/2\big)\Big)+j\phi_n\Bigg) \nonumber
\eea
with
\be
G(\phi_n)=-285v_R^4+366 v_R^2 \phi_n^2-81 \phi_n^4\, ,
\ee
and $\lambda_R,v_R, \bar \Lambda_R$ are the renormalised parameters given in Appendix \ref{static}. 
The specific form (\ref{Sigman}), including the renormalised parameters, is chosen in such a way that, in the absence of source we have
\be
\Sigma_n[0]=\int \text{d}^4 x \sqrt{g}\,  \bar \Lambda_R~,
\ee
which makes the discussion on vacuum energy simpler. Note that, in eq.(\ref{Sigman}), the static saddle points $\phi_n$ can be expressed as in 
eq.(\ref{staticsaddle}), where the parameters can be replaced by the renormalised ones, since they satisfy the equation of motion \cite{Alexandre:2022sho}.

\subsection{Instanton gas}\label{Sec.gas}

\begin{figure}[h!]
     \centering
     \includegraphics[width=0.5\textwidth]{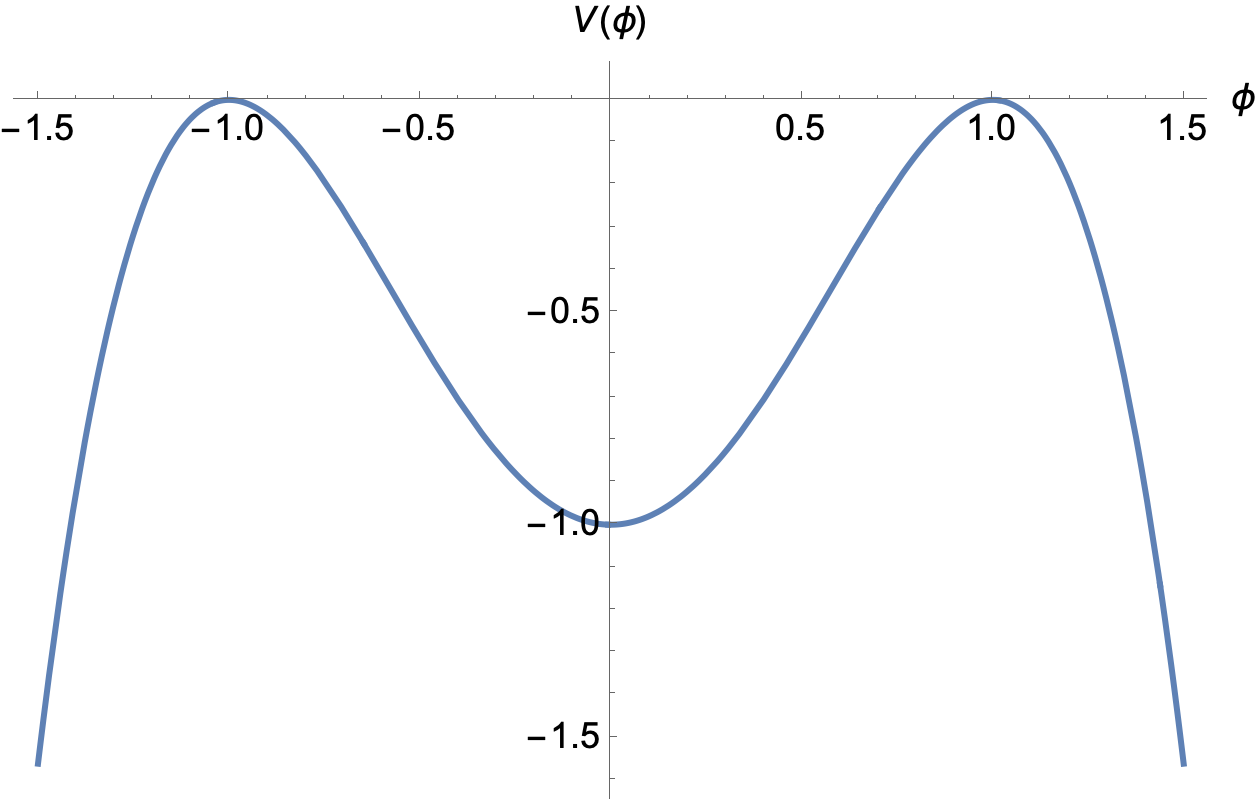}
    \caption{The upside-down potential $V(\phi)$ in which the field oscillates. 
    One instanton corresponds to the motion from infinitesimally close to one hilltop to infinitesimally close to the other.}
    \label{upsidedown}
\end{figure}

We describe here Euclidean time-dependent saddle points. 
In the absence of a source, they obey the following equation
\be \label{eq:eom-instanton}
\ddot\phi+\omega^2 \phi -\frac{\lambda}{6} \phi^3=0~,
\ee
where $\omega=v\sqrt{\lambda/6}$, which corresponds to a problem of real-time classical mechanics in the 
upside-down potential 
\be
V(\phi)=-\frac{\lambda}{24}(\phi^2-v^2)^2~,
\ee
represented in Fig. \ref{upsidedown}. 

\begin{figure}[h!]
     \centering
     \begin{subfigure}[b]{0.5\textwidth}
         \centering
         \includegraphics[width=\textwidth]{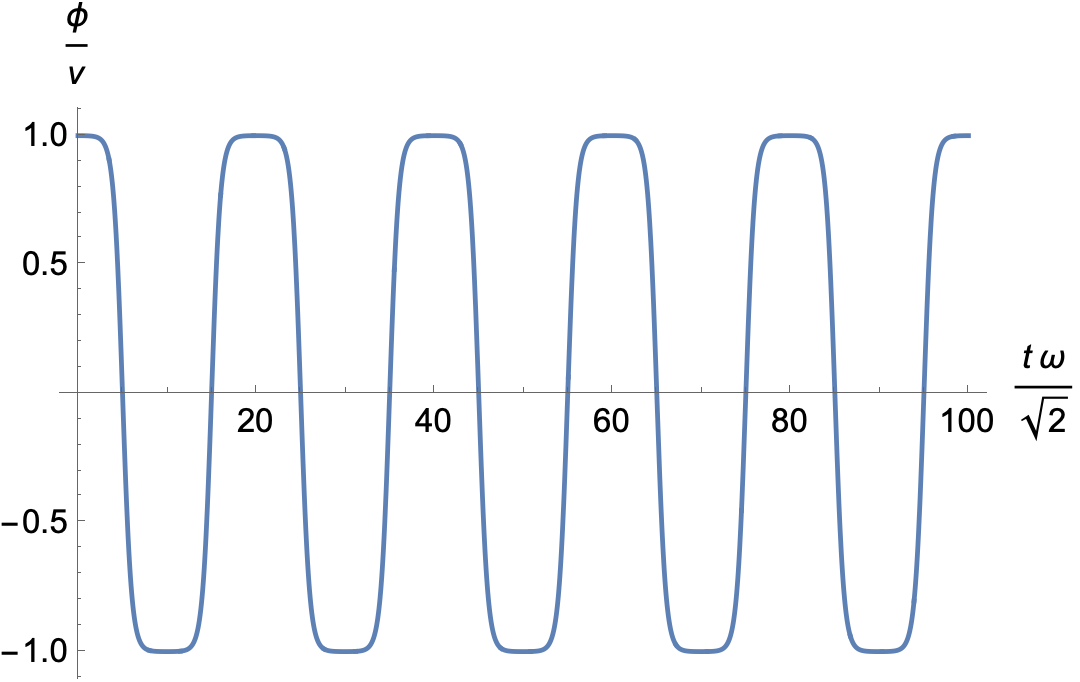}
         \caption{An exact saddle point configuration, corresponding to periodic oscillations between the two hills provided by 
         the upside-down potential shown on Fig.\ref{upsidedown}.}
         \label{exactinst}
     \end{subfigure}
     \hfill
     \begin{subfigure}[b]{0.5\textwidth}
         \centering
         \includegraphics[width=\textwidth]{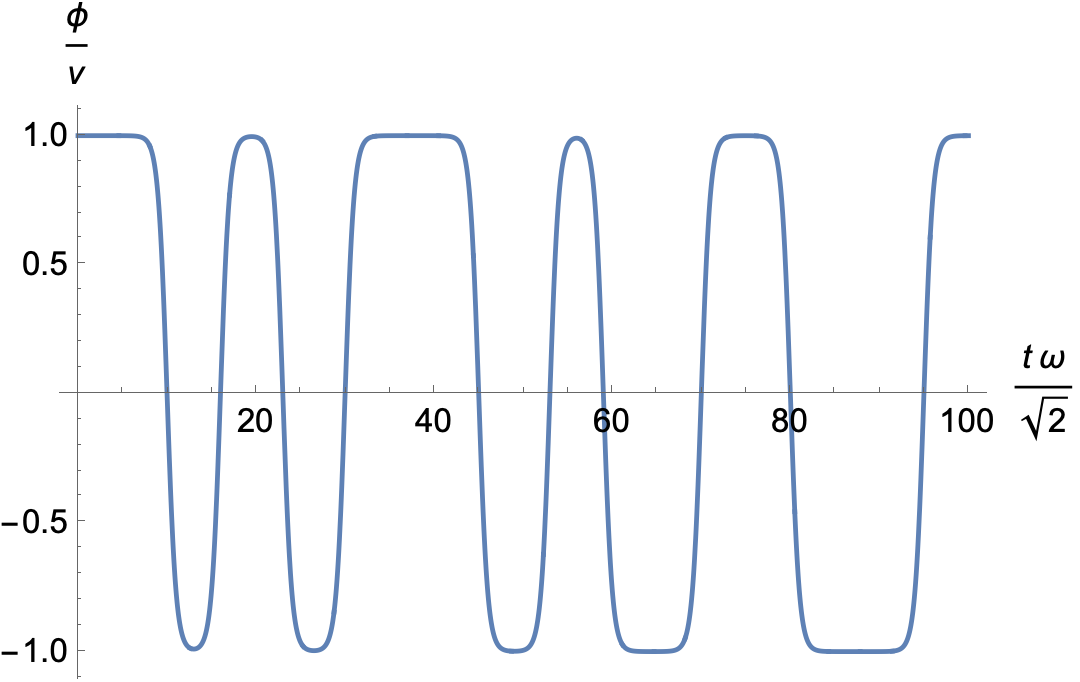}
         \caption{An approximate saddle point configuration with the same number of oscillations, but not periodic.
         The jumps are randomly distributed, but the average distance between them is larger than their width, 
         such that they keep their shape and the action of the configuration is essentially the same as the action 
         for the exact saddle point in fig.(a).}
         \label{approxinst}
     \end{subfigure}
        \caption{Example of exact and approximate saddle points. In the dilute gas approximation, the difference between the corresponding actions 
        is exponentially small, and the partition function is dominated by the whole set of approximate saddle points.}
\end{figure}

The motion starting asymptotically close to a hilltop and ending asymptotically close to the other hilltop is given by the known solution
\be
\phi_{inst}(j=0)=\pm v\tanh\left(\frac{\omega}{\sqrt{2}}(t-t_1)\right)~,
\ee
where $t_1$ corresponds to the ``jump", where the instanton goes through 0, 
and the corresponding action is
\be\label{Sinstj=0}
S[\phi_{inst}(j=0),\bar \Lambda=0]=a^3 S_0~~~~\mbox{with}~~~~S_0\equiv \frac{4\sqrt2}{\lambda}\omega^3V_0~.
\ee
Indeed, the field spends a large (Euclidean) time close to a hilltop, with an exponentially small 
contribution to both the potential and the kinetic energy, and the main contribution to the action comes from the jump.
For $p$ jumps, an exact saddle point is a series of periodic oscillations between the two hills.
If the motion starts exponentially close to a hilltop, the distance $|t_{i+1}-t_i|$ between two consecutive jumps is large compared to 
the width $2\pi/\omega$ of a jump. The motion is then approximately described by 
\be
\phi_{inst}^{(p)}(j=0)\simeq~ \pm v\tanh\left(\frac{\omega}{\sqrt{2}}(t-t_1)\right)
\times\tanh\left(\frac{\omega}{\sqrt{2}}(t-t_2)\right)\times\cdots
\times\tanh\left(\frac{\omega}{\sqrt{2}}(t-t_p)\right)~,
\ee
where the times $t_i$ are regularly spread along the Euclidean time (see Fig.\ref{exactinst}), and the corresponding action is
\be
S[\phi_{inst}^{(p)}(j=0),\bar \Lambda=0]\simeq p~ a^3  S_0~.
\ee
The above action remains unchanged when the jumps are shifted though, provided the condition
$|t_{i+1}-t_i|\gg2\pi/\omega$ is satisfied, which is called the dilute gas approximation (see Fig.\ref{approxinst}). 
As a consequence, all the corresponding configurations in the path integral $Z$ contribute as much as the exact 
solution of the equation of motion. The invariance of the action under the translation of jumps has a high degeneracy,
making this dilute gas dominant in $Z$.

We show in Appendix \ref{instanton} that,
in the presence of a source, the summation over all the $p$-jump saddle points leads to
\be
\Sigma_{gas}[j]\simeq\frac{1}{2}\big(\Sigma_1[j]+\Sigma_2[j]\big)-\hbar\ln\big(\exp(\bar N)-1\big)
\equiv-\hbar\ln Z_{gas}[j]~,
\ee
where the statistical average number of jumps between the two static saddle points is 
\be\label{Nbar}
\bar N=\sqrt{g_{00}}~\omega T\sqrt\frac{6a^3S_0}{\hbar\pi} e^{-a^3S_0/\hbar}~.
\ee
We note that the parameters in the latter equation can be understood as the renormalised ones, since the contribution of $\bar N$ is at one-loop already.

The exponential of $\bar N$ appearing in the partition function is a known feature in tunnelling studies, and it arises from the summation over the zero modes of each (anti-) instanton (see Appendix \ref{instanton} for details). Note that we are interested here in the situation where $S_0$ is fixed and the total Euclidean time $T$ goes to infinity, such that $\bar N$ is assumed to be large. An alternative situation, relevant at finite temperature, consists in fixing $T$ and taking $S_0\to\infty$, such that $\bar N\to0$. This corresponds to the suppression of tunnelling, and where SSB provides a better description of the system \cite{Alexandre:2022qxc}. The expression (\ref{Nbar}) can be understood as the total Euclidean time $T$ multiplied by the tunnelling rate $\omega\sqrt{6a^3S_0/\hbar\pi}~e^{-a^3S_0/\hbar}$.

\section{Effective action}\label{Sec.effectiveaction}

We describe here the main steps for the construction of the effective theory, as well as its energetic properties. The details can be found in Appendix \eqref{ap:effective}.

\subsection{Symmetric ground state}

From the previous section, the partition function can be expressed as 
\bea
Z[j]&\simeq&Z_1[j]+Z_2[j]+Z_{gas}[j]\\
&=&\exp\left(-\frac{1}{\hbar}\Sigma_1[j]\right) + \exp\left(-\frac{1}{\hbar}\Sigma_2[j]\right) 
+ (\exp(\bar N)-1) \exp\left(-\frac{1}{2\hbar}\Big(\Sigma_2[j]+\Sigma_2[j]\Big)\right)~,\nonumber
\eea
from which one can derive the classical field $\phi_c$, which corresponds to the vacuum expectation value in the presence of the source $j$
\be \label{eq:phic-j}
\phi_c=\frac{-\hbar}{Z(j)\sqrt{g}}\frac{\delta Z}{\delta j}=-M^{-2} \, j +{\cal O}(j^3)~ .
\ee
In the previous expression and in the limit where $T\to\infty$, we show in Appendix \ref{ap:effective} that
\be\label{M^-2}
M^{-2}=\frac{3}{\lambda_R v_R^2}\left(1+\frac{27\hbar\lambda_R}{32\pi^2}\right)+\mathcal{O}(\hbar^2)~.
\ee
We note that $\phi_c$ is proportional to $j$, showing symmetry restoration: the vacuum for $j=0$ is at $\phi_c=0$.\\
The relation $\phi_c[j]$ is then inverted to 
\be
j[\phi_c]=-M^2\phi_c+{\cal O}(\phi_c^3)~,
\ee
and the 1PI effective action, defined through the Legendre transform as a functional of $\phi_c$, is
\bea\label{Gammafinal}
\Gamma[\phi_c]&=&-\hbar\ln Z[j\big[\phi_c]\big]-\int \text{d}^4x\sqrt{g}~\phi_c ~ j[\phi_c]\\
&=&\Gamma[0]+\frac{1}{2}\int \text{d}^4 x \sqrt{g}~M^2\phi_c^2+\mathcal{O}(\phi_c^4)~.\nonumber
\eea
In the previous expression, the effective action for the ground state reads
\bea \label{eq:EE-ground}
\Gamma[0]&=& \int \text{d}^4x\sqrt{g}\, \bar \Lambda_R - \hbar \ln(e^{\bar N}+1)\\
&\simeq&\int \text{d}^4x\sqrt{g}\, \bar \Lambda_R - \hbar\bar N~.\nonumber
\eea
To summarise the essential features of the effective action (\ref{Gammafinal}):
\begin{itemize}
\item it is convex, since $M^2>0$, and has its ground state at $\phi_c=0$;
\item the ground state energy has a non-trivial dependence on the comoving volume, via $\bar N$, 
and is therefore not extensive in the usual thermodynamical sense.
\end{itemize}

\subsection{NEC violation}

For simplicity, in what follows we will drop the sub-index ${}_R$ and all the parameters should be understood as the renormalised ones.

We focus here on the fluid provided by the ground state $\phi_c=0$. 
In order to obtain the energy density and the pressure, we need to represent $\Gamma[0]$ and thus $\bar N$ as the integral over a Lagrangian density, restoring the time dependence of the scale factor. This is done in
 Appendix \ref{ap:effective}, where we show that the expression (\ref{eq:EE-ground}) can be written as
\bea\label{eq:effective-final-0}
\Gamma[0]=\int \text{d}^4x\sqrt{g}\left( \bar \Lambda-\rho_0~\frac{e^{-\alpha^3}}{\alpha^{3/2}}\right)~,
\eea
where 
\be\label{rho0}
\alpha^3\equiv a^3 \frac{S_0}{\hbar}~~~~\mbox{and}~~~~ \rho_0\equiv\hbar\, \frac{\omega}{V_0}\frac{S_0}{\hbar}\sqrt\frac{6}{\pi}=\frac{2\,\lambda~ v^4}{3\sqrt{3\pi}}~ .
\ee
From eq.\eqref{eq:effective-final-0}, the energy density and the pressure are obtained from the components of the energy-momentum tensor
\be
T_{\mu\nu}=\frac{2}{\sqrt{g}}\frac{\delta\Gamma[0]}{\delta g^{\mu\nu}}~,
\ee
and we find 
\bea \label{rhop}
\rho&=&\bar \Lambda-\rho_0~\frac{e^{-\alpha^3}}{\alpha^{3/2}}~,\\
p&=&-\bar \Lambda+\rho_0\left(\frac{1}{2\alpha^{3/2}}-\alpha^{3/2}\right)e^{-\alpha^3}~.\nonumber
\eea
The fluid provided by the ground state therefore features the following properties:
\begin{itemize}
\item It consistently satisfies the (real-time) continuity equation $\dot\rho+3H(\rho+p)=0$;
\item It violates the NEC  
\be
\rho+p=-\rho_0~e^{-\alpha^3}\left(\alpha^{3/2}+\frac{1}{2\alpha^{3/2}}\right)~<0~;
\ee
\item  It can induce a cosmic bounce since it can simultaneously satisfy $\rho=0$ and $\rho+3p<0$. 
Assuming $\bar \Lambda > 0$, the second condition is automatically satisfied if the NEC is violated;
\item Assuming $e^{-\alpha^3}\ll1$, its equation of state has the phantom form
\be 
w=\frac{p}{\rho}\simeq-1-\frac{\rho_0}{\bar\Lambda}\,e^{-\alpha^3}\left(\alpha^{3/2}+\frac{1}{2\alpha^{3/2}}\right)~<-1~.
\ee
\end{itemize}
We stress here an important point: the property $w<-1$ does not arise from a kinetic energy with the opposite sign, but is a consequence of tunnelling between the two degenerate bare vacua, which induces a homogeneous symmetric ground state.

\section{Friedmann Equations}\label{Sec.Friedmann}

\begin{figure}[h] 
     \centering
     \begin{subfigure}[b]{0.5\textwidth}
         \centering
         \includegraphics[width=\textwidth]{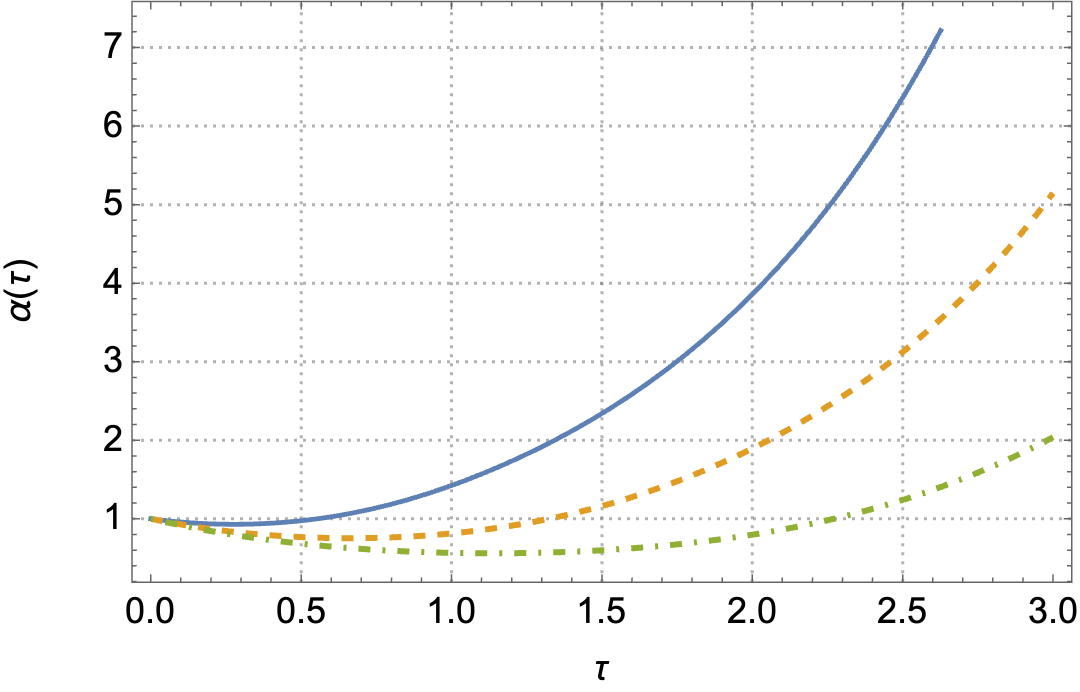}
     \end{subfigure}
     
     \hfill
     
     \begin{subfigure}[b]{0.5\textwidth}
         \centering
        \includegraphics[width=1.01\textwidth]{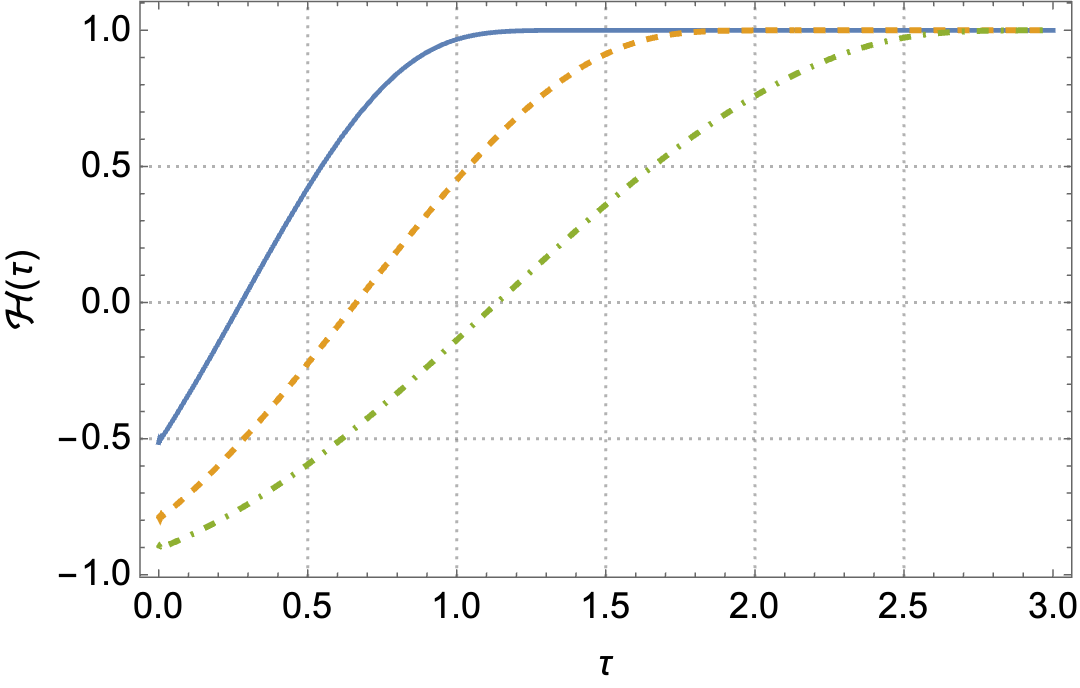}  
     
     \end{subfigure}
        \caption{Time evolution of the scaled scale factor $\alpha$ (upper panel) and the scaled Hubble rate $\mathcal{H}=\alpha'/\alpha$ (lower panel) with initial condition $\alpha(0)=1$, 
        for three different values of $r$, namely $r=2$ (solid line), $r=1$ (dashed line) and $r=0.5$ (dashed-dotted line). }
        \label{fig:scale-factor}
\end{figure}

In this section we go back to Lorentzian signature. As explained in the introduction, we study the back-reaction of the effective theory on the metric,
such that the energy-momentum tensor in the Einstein equations $G_{\mu\nu}=\kappa T_{\mu\nu}$ contains the energy density and pressure given by eqs. (\ref{rhop}), 
and $\kappa$ is the renormalised gravity coupling. The resulting Friedmann equations read
\bea\label{Friedman}
H^2&=&\frac{\kappa}{3}\rho\\
\frac{\ddot a}{a}&=&-\frac{\kappa}{6}(\rho+3p)~,\nonumber
\eea
that we study here numerically. The first equation $H^2\propto\rho$ gives the initial condition $\dot a_0$ once $a_0$ is known,
and the second equation provides the evolution equation for $a(t)$. We then introduce the dimensionless time
\be
\tau\equiv t~\sqrt\frac{\Lambda}{3}~,
\ee
and we use the expressions (\ref{rhop}) to obtain from eqs.(\ref{Friedman})
\bea \label{eq:friedman-rescaled}
\alpha'&=&\pm\alpha\sqrt{1-r~\frac{e^{-\alpha^3}}{\alpha^{3/2}}}\\
\frac{\alpha''}{\alpha}&=&1-\frac{r~e^{-\alpha^3}}{4~\alpha^{3/2}}(1-6\alpha^3)~,\nonumber
\eea
where a prime denotes a derivative with respect to $\tau$ and 
\be\label{r}
r=\kappa\frac{\rho_0}{\Lambda}=\frac{\rho_0}{\bar\Lambda}~.
\ee
The Friedmann Equations \eqref{eq:friedman-rescaled} are solved numerically, and we plot in Figure \ref{fig:scale-factor} the solutions corresponding to a fixed value of $\alpha(0)$ and different values of the parameter $r$. The initial condition for $\alpha'(0)$ is given by the negative branch $\alpha'(0)<0$ of the first Friedmann equation, 
in order to describe a cosmological bounce, which will occur if $\rho=0$ and $\rho+3p<0$ are simultaneously satisfied at a given time $t_b$. We see that such a bounce is indeed 
generated, after which the expansion suppresses tunnelling:  the NEC is recovered and the metric dynamics enters a de Sitter phase, with constant $H$.

\section{Conclusions}

We have described how the energetic properties arising from tunnelling could be relevant in a cosmological context, starting from standard QFT and Einstein gravity. 
To summarise the non-perturbative mechanism described in this article:
\begin{enumerate}[{\it (a)}]
\item  The effective theory taking into account tunnelling between two degenerate vacua is obtained by considering the contribution of different saddle points in the partition function; 
\item  As a consequence of this interplay between the two vacua $\pm v$, the resulting true vacuum is at $\phi_c=0$, with an energy which is not proportional to the comoving volume;
\item  This non-extensive feature of the vacuum energy implies NEC violation; 
\item  The NEC violation induces a cosmological bounce in the case of initial spacetime contraction, and is valid until the resulting expansion suppresses tunnelling, such the ANEC is satisfied.
\end{enumerate}
The adiabatic approximation is well justified in the vicinity of the cosmological bounce, but out-of-equilibrium studies would be necessary to include the full-time dependence of the scale factor if one wishes to look at what happens away from the bounce. A related improvement to this work would be to derive our results in a manifestly covariant way.

Regarding the assumption of finite-volume FLRW space-time, this study has required
a toy-model geometry/topology, in the form of a 3-torus or 3-sphere, and thus
still needs to be developed for phenomenological purposes.
Also, quantum corrections in a finite volume should in principle take into account discrete momentum, as well as periodic boundary conditions. This is done in the 
framework of Casimir effect studies \cite{Bordag:2001qi}, whereas the present article focuses on the tunnel effect, with continuous momentum and effectively 
Dirichlet boundary conditions. A natural step further would then consider a discrete spectrum, which could be done numerically for example. 

The situation of non-degenerate minima would avoid making the finite-volume assumption, 
since the relevant instanton action (the Coleman bounce saddle point) is independent of the volume. In this case, quantum fluctuations for the latter saddle point would involve an imaginary part, which should be cancelled by the imaginary part induced by other saddle points \cite{Andreassen:2016cvx}, since the effective potential is real. The whole process is challenging to describe analytically in more than 0-dimensional space-time though, but is a potential avenue to explore, since it could be relevant as a component of Dark Energy
(see related comment below). There are questions which remain to be answered in the situation of non-degenerate minima though, 
one of which is the possibility to obtain a NEC-violating effective ground state.
Indeed, if one considers only $O(4)$-symmetric saddle points, the NEC should not be violated since space and (Euclidean) time are treated in the same way. A more complex saddle point structure, 
to be studied in the context of real-time tunnelling \cite{Ai:2019fri}, could involve NEC violation though, but this study goes beyond the scope of the present article.
Another question in this context is the use of a homogeneous classical field, which would be valid at large scales only (large compared to the radius of a 4-dimensional bubble), 
in order to describe an average over the gas of bubbles. But strictly speaking, the homogenous configurations used in the present work are valid for a second-order quantum phase transition only,
whereas the presence of bubbles is a signature for a first order phase transition.

In our model, the universe asymptotically approaches a de Sitter phase in the late-time regime, from which one cannot escape.
This phase could potentially represent Dark Energy in the late Universe, where
we would need to identify $\Lambda$ with the current cosmological constant. Given our numerical study where
$r\sim1$, and assuming $S_0\sim\hbar$ for tunnelling to be significant, we find 
from the expressions (\ref{Sinstj=0}), (\ref{rho0}) and (\ref{r}) for $S_0$, $\rho_0$ and $r$ respectively, that
\be
\Lambda\sim\hbar\sqrt\frac{\lambda}{\pi}\frac{\kappa v}{V_0}~.
\ee
The latter order of magnitude allows an estimate of the volume $V_0$ at the bounce, which should be much smaller than
the volume $V$ of the current visible Universe, for tunnelling to be significant enough. $V_0$ should 
also be much larger than the Plank volume $V_P$, for the classical gravity regime to be valid.
We obtain then the constraint
\be
V_P\ll \hbar\sqrt\frac{\lambda}{\pi}~\frac{\kappa v}{\Lambda}\ll V~,
\ee
which could be helpful in model building for a bouncing Cosmology scenario. In any case, one should be careful when drawing conclusions
far from the bounce, since our approach is based on equilibrium QFT, which is valid only in the vicinity of the bounce, i.e. when 
the condition (\ref{validity}) is satisfied. A more complete study would involve the Keldysh formalism for example, allowing out-of-equilibrium
predictions, and which is a potential next step.

\section*{Acknowledgements}
 The authors thank Katy Clough and Malcolm Fairbairn for cosmology-related discussions, 
as well as Jose Navarro-Salas and Janos Polonyi for insightful comments.
This work is supported by the Leverhulme Trust (grant RPG-2021-299) and the Science and Technology
Facilities Council (grant STFC-ST/T000759/1). For the purpose of Open Access, 
the authors have applied a CC BY public copyright licence to any Author Accepted Manuscript version arising from this submission.

\appendix

\section{One-loop effective action in curved space-times} \label{static}

In this appendix we review the main steps to obtain the one-loop effective action in curved space-times for a real scalar field 
in a double-well potential, and propagating in a curved background with Euclidean signature. 
We focus here on one saddle point only.

For renormalisation purposes, we need to consider the bare action of this model 
\be \label{eq:Lagrangian-1b}
S[\phi,g]=\int \text{d}^d x \sqrt{g}\left(\frac{1}{2}g^{\mu \nu}\partial_\mu \phi \partial_\nu \phi + \frac{1}{2}\xi R \phi^2
+\frac{\lambda}{4!}(\phi^2-v^2)^2+\bar\Lambda+j \phi\right)~,
\ee 
together with the semi-classical action for gravity\footnote{ We note that the Euclidean form of the Lagrangian differs with a minus sign with respect to its Lorentzian form.}
\be \label{eq:action-semi-grav}
S_{G}[g]=-\int \text{d}^d x\sqrt{g}\Big[(2 \kappa)^{-1} R 
+(\epsilon_1 R^2 + \epsilon_2 R^{\mu \nu}R_{\mu \nu}+\epsilon_3 R^{\mu \nu \rho \sigma}R_{\mu \nu \rho \sigma})\Big]~,
\ee
in $d$ space-time dimensions, where $\kappa=8 \pi G$, and $\bar \Lambda=\kappa^{-1} \Lambda$. For convenience, we have included the cosmological constant term in the matter sector. The inclusion of the higher curvature terms is needed for the cancellation of the divergences that arise in this context. In this setup, the Klein-Gordon equation for the scalar field is
\be
(-\Box_E+\xi R-\tfrac{\lambda}{6}v^2  +\tfrac{\lambda}{3!} \phi^2)\phi +j=0~,
\ee
where $\Box_E=g^{\mu\nu}\nabla_\mu \nabla_{\nu}=\frac{1}{\sqrt{g}}\partial_\mu(\sqrt{g} g^{\mu\nu}\partial_\mu)$,
and the scalar field can be expanded around a saddle point $\phi=\phi_s + \delta \phi$.
The associated Euclidean Green's function for the quantum fluctuation $\delta \phi$ reads
\be
(-\Box_E+Q) G_E(x,x')=\frac{1}{\sqrt{g}}\delta^{(4)}(x-x')\, ,
\ee
where
\be
Q=\frac{\lambda}{2}\phi_s^2-\frac{\lambda v^2}{6}+\xi R\, .
\ee
The one-loop correction to the classical action can be written in terms of the Green's function as \cite{parker-toms}
\bea \label{eq:effective-total-grav}
\Sigma[\phi_s,g]&=&S_G[g]+S[\phi_s,g]-\tfrac{1}{2}\hbar\, \textrm{ln}\,  \textrm{Det}\, G_E\,\\
&\equiv& S_G[g]+S[\phi_s,g] +\Sigma^{(1)}[\phi_s,g]\, . \nonumber 
\eea
For general background configurations, the Green's function is unknown. However, an approximated expression for the  quantum contribution $\Sigma^{(1)}[\phi_s,g]$ in the case of slowly varying background fields  $\phi_s$ and $g$ 
can be computed using the proper-time formalism as follows (see Refs. \cite{Hu-lambdaphi4,Rajantie18} for a detailed explanation). 

The DeWitt-Schwinger representation of the propagator $G_E(x,x')$ is given by
\be
G_E(x,x')=\int_0^{\infty} \text{d}s\,  H(x,x';s)\, ,
\ee
where the kernel $H(x,x';s)$ obeys a diffusion equation with appropriate boundary conditions \cite{Vassilevich03}. 
For the one-loop connected graph, it translates into 
\be
\Sigma^{(1)}[\phi_s,g]=\, \frac{\hbar}{2}\int \text{d}^d x \sqrt{g}\int_0^{\infty} \frac{\text{d}s}{s} \,  H(x,x;s)\, .
\ee
The kernel $H(x,x';s)$ admits, in general, an asymptotic expansion in terms of the Schwinger proper-time parameter \cite{Schwinger51}. At coincidence $x'\to x$ this expansion reads 
\be \label{eq:heatK}
H(x,x;s)=\frac{e^{-m^2 s}}{(4 \pi  s)^{d/2}}\sum_{k=0}^{\infty}a_k(x)\, s^k\, .
\ee
where $a_k(x)$ are the so-called the deWitt coefficients  and $d$ is the space-time dimension. The first few coefficients are \cite{Gilkey75,Vassilevich03} 
\bea
a_0&=&1\, ;\\
a_1&=&\frac{1}{6}R-Q\, ;\\
a_2&=&-\frac{1}{180} R_{\alpha \beta \gamma \delta} R^{\alpha \beta \gamma \delta}-\frac{1}{180} R^{\alpha \beta} R_{\alpha \beta}
-\frac{1}{30}\Box_E R\\
&&+\frac{1}{6}\Box_E Q + \frac{1}{2}Q^2-\frac{1}{6}R Q +\frac{1}{72} R^2\, . \nonumber
\eea
This expansion captures, in its leading orders, the UV divergences ($s\to 0$) of the theory and it is routinely used for renormalisation in the context of QFT in curved spaces.  

The expansion above \eqref{eq:heatK} has an important property: it admits an exact resummation \cite{R-summed1,R-summed2}
\be
H(x,x;s)=\frac{e^{-\mathcal{M}^2 s}}{(4 \pi  s)^{d/2}}\sum_{k=0}^{\infty}b_k(x)\, s^k\, ,
\ee
with 
\be
\mathcal M^2= Q-\frac{1}{6}R\, ,
\ee
such that, the new coefficients $b_k(x)$ do not contain any term that vanish when $Q$ and $R$ are replaced by zero. For example, for the first resummed deWitt coefficients we have
\bea
b_0&=&1\,; \\
b_1&=&0\, ;\\
b_2&=&-\frac{1}{180} R_{\alpha \beta \gamma \delta} R^{\alpha \beta \gamma \delta}-\frac{1}{180} R^{\alpha \beta} R_{\alpha \beta}
-\frac{1}{30}\Box_E R+\frac{1}{6}\Box_E Q~ .
\eea
Therefore, the resummed expansion becomes a derivative expansion in the field $\phi_s$ and the metric, physically meaningful in the case of slowly varying background fields. 
Then, it is possible to truncate the expansion at a given order $N$ - the order of derivatives - to obtain an approximated expression for the  one-loop connected graph
\be
\Sigma^{(1)}[\phi_s,g]=\, \frac{\hbar }{2}\int \text{d}^d x \sqrt{g}\int_0^{\infty} \frac{\text{d}s}{s} ~~  
\frac{e^{-\mathcal{M}^2 s}}{(4 \pi  s)^{d/2}}\sum_{k=0}^{N}b_k(x)\, s^k\,  .
\ee
The expression above is divergent for $d=4$ and can be renormalised using dimensional regularization. 
For arbitrary dimension $d$, the proper-time integrals can be performed to give
\be \label{eq}
\Sigma^{(1)}[\phi_s,g]=\frac{\hbar}{(4 \pi)^{d/2}}\left(\frac{\mathcal M}{\mu_d}\right)^{d-4}\int \text{d}^dx \sqrt{g}~  
\sum_{k=0}^{N} b_k(x)  \mathcal{M}^{d-2k}\, \Gamma\left(k-\frac{d}{2}\right) ~.
\ee
We have introduced a renormalisation mass parameter to proceed with dimensional regularization in what follows. Truncating the sum at $N=2$ and expanding around $d\to 4$ we find
\be \label{eq:effective-sigma1}
\Sigma^{(1)}=\hbar\int \text{d}^4 x \sqrt{g} \left[ \frac{ \mathcal{M}^4}{64 \pi^2}\Big[\ln\left(\frac{\mathcal{M}^2}{\mu^2}\right)-\frac{3}{2}\Big]
+ \frac{b_2 }{32 \pi^2} \ln\left(\frac{\mathcal{M}^2}{\mu^2}\right)\right]~,
\ee
 where  $\mathcal{M}^2>0$ since we quantise about stable saddle points and the curvature effects are expected to be small. In the above expression, the divergences have been absorbed in the scale parameter $\mu$, which is defined by 
\be
\ln \mu^2=\ln\Big(4 \pi \mu_d^2\Big)-\gamma-\frac{2}{d-4}~~~(\mbox{finite when}~d\to4)~.
\ee

From these expressions, we can directly obtain the renormalised values of the coupling constants of the problem (see, for example Ref. \cite{Hu-book}). 

In our particular problem, we are assuming an adiabatic expansion of the universe, and that quantum processes under consideration occur at equilibrium. 
Hence, we neglect the curvature of space-time. Hence we are only interested in the couplings $\lambda,~v,~\bar \Lambda$.
For simplicity, we will follow \cite{Markkanen13}, and apply the renormalisation conditions at the same scale for all bare parameters, namely,
\bea
3\frac{\partial^2 L}{\partial \phi^2}\Big|_{\phi=\pm v_R, g=\eta}= \lambda_R\, v_R^2\, , \qquad  \frac{\partial^2 L}{\partial \phi^4}\Big|_{\phi=\pm v_R, g=\eta}= \lambda_R\, , \qquad 
 + L\Big|_{\phi=\pm v_R, g=\eta}= \bar \Lambda_R ~, 
\eea
where $\eta$ is the Euclidean flat metric and $\Sigma=\int \text{d}^4x\sqrt{g}~L$. From these conditions we obtain 
\bea
\delta \lambda&=&\frac{3 \lambda_R^2}{32 \pi^2}\left(3+\ln(\frac{v_R^2\lambda_R}{3\mu^2})\right)\, ,\\
\delta v^2 &=& \frac{v_R^2 \lambda_R}{16 \pi^2}\left(10-\ln(\frac{v_R^2\lambda_R}{3\mu^2})\right)\, ,\\
\delta \bar \Lambda&=& \frac{v_R^4 \lambda_R^2}{1152 \pi^2}\left(-3+2\ln(\frac{v_R^2\lambda_R}{3\mu^2})\right)~, \label{eq:delta-Lambda}
\eea
where we define $\lambda_R=\lambda+\hbar ~\delta\lambda$, $v_R^2=v^2+\hbar~\delta v^2$,
and  $\bar \Lambda_R=\bar \Lambda+\hbar ~\delta \bar \Lambda$.  Inserting these results in \eqref{eq:effective-total-grav} and assuming $R=0$ and $\phi_s$ static, we obtain the final renormalised connected graph given in Sec. \ref{subsec:static-s}.

For completeness, we also give the renormalised values of $\kappa^{-1}$ and $\xi$. The renormalisation conditons we impose are
\be
-2\frac{\partial L}{\partial R}-\xi_R \phi^2 \Big|_{\phi=\pm v_R,g=\eta}=\kappa^{-1}_R\, , \qquad \frac{\partial^3 L}{\partial R \partial \phi^2}\Big|_{\phi=\pm v_R,g=\eta}=\xi_R\, ,
\ee
that lead to
\bea
\delta \xi&=&\frac{\lambda_R(6\xi_R-1)}{192\pi^2}\left(3+\ln(\frac{v_R^2\lambda_R}{3\mu^2})\right)\, ,\\
\delta(\kappa^{-1})&=&\frac{v_R^2 \lambda_R(6\xi_R-1)}{2304 \pi^2}\left(11+\ln(\frac{v_R^2\lambda_R}{3\mu^2})\right)\,.
\eea

\section{Quantisation over instanton configurations}\label{instanton}

In Section \ref{Sec.gas} we describe few features of the gas of instantons for a vanishing source. 
In the presence of an infinitesimal source $j\ll j_c$, the jump is not modified, 
and what changes is the position of the 
asymptotically "flat" parts of the instantons, which now go from one saddle point $\phi_i(j)$ to the other,
instead of going from one vacumm $\pm v$ to the other $\mp v$. We have then, instead of eq.(\ref{Sinstj=0}),
\be
S[\phi_{inst}(j)]\simeq a^3  S_0+\frac{1}{2}\big(S_1[j]+S_2[j]\big)~,
\ee
since on average the configuration spends half the Euclidean time exponentially close to 
$\phi_1(j)$ and the other half close to $\phi_2(j)$. The contribution of one instanton $F_{inst}\exp(-S[\phi_{inst}]/\hbar)$ to the partition function
is the product of the following contributions
\begin{itemize}
    \item The "flat" part close to each static saddle point, leading to the fluctuation factor $F_i$ about each
    of the static saddle points, for half of the total Euclidean time  
    \be
    \sqrt{F_1F_2}e^{-(S_1+S_2)/(2\hbar)}=\exp\left(-\frac{1}{2\hbar}\big(\Sigma_1[j]+\Sigma_2[j]\big)\right)~,
    \ee
    where $\Sigma_n[j]$ is given in eq.(\ref{Sigman}).
    \item Fluctuations above one jump which, discounting the zero mode corresponding to the 
    translational invariance of the jump, lead to the factor (see \cite{Kleinert:2004ev,Dunne2009})
    \be
    \sqrt\frac{6a^3S_0}{\hbar\pi}~; 
    \ee
    
    \item The zero mode corresponding to the position of the jump, 
    which can happen at any Euclidean time between 0 and $T$, and thus gives the extra factor
    \be
    \omega\int_0^T \sqrt{g_{00}}~\text{d}t=\sqrt{g_{00}}~\omega T~.
    \ee
    Note that the summation over the different positions of the jump is done with the comoving proper time, 
    since the jump is observed by the comoving observer. Here, $S_0$ and $\omega$ are defined with the renormalised parameters.
\end{itemize}
All together, the contribution of one instanton to the partition function is
\be
F_{inst}\exp\left(-\frac{S[\phi_{inst}]}{\hbar}\right)=\sqrt{g_{00}}~\omega T\sqrt\frac{6a^3S_0}{\hbar\pi}
\exp\left(-a^3\frac{S_0}{\hbar}-\frac{1}{2\hbar}\big(\Sigma_1[j]+\Sigma_2[j]\big)\right)~.
\ee
For a $p$-jump saddle point in the dilute gas approximation, and where
the width of an instanton is negligible compared to the total Euclidean time $T$, the classical action is
\be
S[\phi_{inst}^p(j)]\simeq pa^3  S_0+\frac{1}{2}\big(S_1[j]+S_2[j]\big)~.
\ee
Also, whereas the first jump can happen at any time $t_1\in[0,T]$, 
the jump $i$ can happen at a time $t_i\in[t_{i-1},T]$ only,  
such that the degeneracy of a $p$-jump configuration leads to the factor \cite{Kleinert:2004ev}
\be
\prod_{i=1}^p\left(\omega\int_{t_{i-1}}^T \sqrt{g_{00}}~\text{d}t_i\right)
=\frac{1}{p!}(\sqrt{g_{00}}~\omega T)^p~~~~~~~(\text{with}~t_0=0)~.
\ee
Summing over all the possibilities for $p$, we obtain the final expression for the 
dilute gas contribution to the partition function
\bea
\exp\left(-\frac{1}{\hbar}\Sigma_{gas}[j]\right)&=&\sum_{p=1}^\infty \frac{1}{p!}(\sqrt{g_{00}}~\omega T)^p
\left(\frac{6a^3S_0}{\hbar\pi}\right)^{p/2}
\exp\left(-pa^3\frac{S_0}{\hbar}-\frac{1}{2\hbar}\big(\Sigma_1[j]+\Sigma_2[j]\big)\right) \nonumber\\
&=&\exp\left(-\frac{1}{2\hbar}\big(\Sigma_1[j]+\Sigma_2[j]\big)\right)
\left[\exp\left(\sqrt{g_{00}}~\omega T\sqrt\frac{6a^3S_0}{\hbar\pi} e^{-a^3S_0/\hbar}\right)-1\right]~.\nonumber
\eea

\section{Effective action, energy density and pressure}\label{ap:effective}

We give here details on the derivation of the one-loop effective action. We start from the partition function 
\bea
Z[j]&=&Z_1[j]+Z_2[j]+Z_{gas}[j]\\
&=&e^{-\Sigma_1/\hbar} + e^{-\Sigma_2/\hbar} + (e^{\bar N}-1) e^{-(\Sigma_1+\Sigma_2)/2\hbar}~,\nonumber
\eea
where $\Sigma_{2}[j]=\Sigma_1[-j]$ which, for small source, can be expanded as 
\be \label{eq:sigma-expansion-appendix}
\Sigma_{1,2}[j]=\int \text{d}^4x \sqrt{g}\left(\bar \Lambda_R \pm \sigma_{(1)} \, j+ \frac{1}{2} \sigma_{(2)}\,  j^2+\mathcal{O}(j^3)\right)~ ,
\ee
with 
\be
\sigma_{(1)}=v_R-\hbar\, \frac{9\lambda_R v_R }{32 \pi^2}\, , \qquad \sigma_{(2)}=-\frac{3}{v_R^2 \lambda_R}-\hbar\,\frac{81}{32 \pi^2 v_R^2}~ .
\ee
The classical field $\phi_c$ is 
\be 
\phi_c=\frac{-\hbar}{Z(j)\sqrt{g}}\frac{\delta Z}{\delta j} =-M^{-2} \, j +{\cal O}(j^3)~, 
\ee
with 
\be
M^{-2}= -\sigma_{(2)}+\frac{V_{4}}{\hbar} \frac{2}{(e^{\bar N}+1)}\sigma_{(1)}^2=\frac{3}{\lambda_R v_R^2}\left(1+\frac{2A}{3}+\hbar \lambda_R\, \frac{27}{2\pi^2}\left(\frac{1}{16}-A\right) \right)+\mathcal{O}(\hbar^2)\, ,
\ee
and 
\be
V_4=\int \text{d}^4x \sqrt{g}~~~~,~~~~A=\frac{V_4 \, \omega^4_R}{\hbar \lambda_R (e^{\bar N}+1)}~.
\ee  
The relation $\phi_c[j]$ is then inverted to $j[\phi_c]$, in order to define the 1PI effective action as the Legendre transform
\be
\Gamma[\phi_c]=-\hbar\ln Z[j\big[\phi_c]\big]-\int \text{d}^4x\sqrt{g}~\phi_c ~ j[\phi_c]~.
\ee
An expansion in the classical field finally gives
\be
\Gamma[\phi_c]=\Gamma[0]+\int \text{d}^4 x \sqrt{g} ~\frac{M^2}{2} \phi_c^2 +\mathcal{O}(\phi_c^4)~ ,
\ee
with 
\bea
M^{2}&=& \left(-\sigma_{(2)}+\frac{V_4}{\hbar}\frac{2}{e^{\bar N}-1} \sigma_{(1)}^2\right)^{-1}\\
&=&\frac{\lambda_R v_R^2}{3}\left(\frac{1}{1+24 A}-\hbar \lambda_R \frac{27}{32\pi^2}\frac{1-16 A}{(1+24 A)^2}\right)+\mathcal{O}(\hbar^2)~,\nonumber
\eea
and 
\be \label{eq:eff-action-0-appendix}
\Gamma[0]= \int \text{d}^4x\sqrt{g}\, \bar \Lambda_R - \hbar \ln(e^{\bar N}+1) \simeq \int \text{d}^4x\sqrt{g}\, \bar \Lambda_R - \hbar \bar N~.
\ee 
In the limit $T\to\infty$ we obtain then
\be
M^{2}=\frac{\lambda_R v_R^2}{3}\left(1-\hbar \lambda_R \frac{27}{32\pi^2}\right)+\mathcal{O}(\hbar^2)~.
\ee
The next step is the analysis of the energy density and pressure for the ground state. The stress-energy tensor can be obtained from the definition 
\be
T^E_{\mu \nu}=\frac{2}{\sqrt{g}}\frac{ \delta \Gamma(0)}{\delta g^{\mu \nu}}~ .
\ee
where we have explicitly written the super-index $^{E}$ as a reminder that we are working in Euclidean signature. Because of homogeneity and isotropy, the stress-energy tensor  can be decomposed as 
\be
T^E_{\mu \nu}=\text{diag}(-\rho,a^2p,a^2 p, a^2 p)~, 
\ee
so that we directly obtain
\be\label{eq:en-p-definitions}
\rho= -T^E_{00}=-\frac{2}{\sqrt{g}}\frac{ \delta \Gamma(0)}{\delta g^{00}}\Big|_{g^{00}=1}\, \, , \qquad p=g^{11}T_{11}^E =\frac{2}{a^2 \sqrt{g}} \frac{ \delta \Gamma(0)}{\delta g^{11}}\Big|_{g^{11}=a^{-2}}\, .
\ee
In order to express $\bar N$ as a Lagrangian density we restore the time dependence of the scale factor with the replacement
\be
\sqrt{g_{00}}~ T~ f(a) \to \int_0^T \text{d}t ~ \sqrt{g_{00}}~ f(a)
\ee
and we express the cell 3-volume at $t=t_0$ as 
\be
V_0=\int \text{d}^3x~a^3(t_0)=\int \text{d}^3x ~~~~\mbox{if we choose}~~~~a(t_0)=1~.
\ee
The effective action for the ground state for $\omega_R T \gg 1$ [see Eq. \eqref{eq:eff-action-0-appendix}] can then be expressed as
\bea\label{eq:eff-action-0-appendix-simplified}
\Gamma[0]&\simeq&\int \text{d}^4x\sqrt{g}~ \bar \Lambda_R
-\hbar\omega_R\sqrt\frac{6S_0}{\hbar\pi}\int_0^{T} \text{d}t \sqrt{g_{00}}\int \frac{\text{d}^3x}{V_0} a^{3/2}~e^{-a^3S_0/\hbar}\\
&=&\int \text{d}^4x\sqrt{g}\left( \bar \Lambda_R-\rho_0~\frac{e^{-a^3S_0/\hbar}}{\sqrt{a^3S_0/\hbar}}\right)~,\nonumber
\eea
where 
\be
\rho_0\equiv\frac{\omega_R S_0}{V_0}\sqrt\frac{6}{\pi}=\frac{2\lambda_R v_R^4}{3\sqrt{3\pi}}~ ,
\ee
and where $S_0$ is defined with the renormalised parameters.

From  Eqs. \eqref{eq:en-p-definitions} and \eqref{eq:eff-action-0-appendix-simplified} we can easily obtain the energy density and the pressure, namely
\bea\label{rhop-appendix}
\rho=-T^E_{00}=-\left.\frac{2}{\sqrt{g}}\frac{\delta\Gamma(0)}{\delta g^{00}}\right|_{g_{00}=1}
&=& + \bar \Lambda_R-\rho_0~\frac{e^{-a^3S_0/\hbar}}{\sqrt{a^3S_0/\hbar}}\, ,\\
p=g^{11}T^E_{11}=\left.\frac{2}{a^2\sqrt{g}}\frac{\delta\Gamma(0)}{\delta g^{11}}\right|_{g_{11}=a^2}
&=&-\bar \Lambda_R+\rho_0\left(\frac{1}{2\sqrt{a^3S_0/\hbar}}-\sqrt{a^3S_0/\hbar}\right)e^{-a^3S_0/\hbar}~.\nonumber
\eea

\bibliographystyle{JHEP.bst} 
\bibliography{bibliography}

\end{document}